\documentclass[10pt,journal,compsoc]{IEEEtran}
\ifCLASSOPTIONcompsoc
  \usepackage[nocompress]{cite}
\else
  \usepackage{cite}
\fi
\ifCLASSINFOpdf
\else
\fi
\usepackage{pgfplots}
\usepackage{atbegshi}
\usepackage{fancyhdr}
\usepackage{caption}
\usepackage[multiple]{footmisc}
\usepackage{graphicx}
\usepackage{amsmath}
\usepackage{multirow}
\usepackage{amssymb}
\usepackage{subcaption}
\usepackage{booktabs}
\usepackage{xcolor}
\usepackage{siunitx}
\usepackage{comment}
\usetikzlibrary{
    patterns,
}
    \pgfplotsset{
        compat=1.7,
        my ybar legend/.style={
            legend image code/.code={
                \draw [##1] (0cm,-0.6ex) rectangle +(2em,1.5ex);
            },
        },
    }
\newcommand*\emptycirc[1][1ex]{\tikz\draw (0,0) circle (#1);} 
\newcommand*\halfcirc[1][1ex]{%
  \begin{tikzpicture}
  \draw[fill] (0,0)-- (180:#1) arc (180:360:#1) -- cycle ;
  \draw (0,0) circle (#1);
  \end{tikzpicture}}

\begin{document}
%
\title{Cloned Identity Detection in Social-Sensor Clouds based on Incomplete Profiles}
%
%
%
%

\author{Ahmed Alharbi,~\IEEEmembership{}
       Hai Dong,~\IEEEmembership{Senior Member, IEEE,}
      ~Xun Yi~\IEEEmembership{}
      and~Prabath Abeysekara~\IEEEmembership{}
\IEEEcompsocitemizethanks{\IEEEcompsocthanksitem A. Alharbi is with the College of Computer Science and Engineering at Taibah University, Medina, Saudi Arabia.\protect\\
E-mail: atharbi@taibahu.edu.sa 
\IEEEcompsocthanksitem H. Dong and X. Yi are with the School of Computing Technologies, RMIT University, Melbourne, VIC, Australia.\protect\\
E-mails: hai.dong@rmit.edu.au; xun.yi@rmit.edu.au
\IEEEcompsocthanksitem P. Abeysekara is with Hitachi Construction Machinery, Brisbane, Queensland, Australia.\protect\\
E-mail: prabathabeysekara@gmail.com}

\thanks{Manuscript received XX XX, XXXX; revised August XX, XXXX. (Corresponding author: Hai Dong)}}

%
%

\markboth{IEEE Transactions on Services Computing,~Vol.~XX, No.~X, XX~XXXX}%
{Shell \MakeLowercase{\textit{et al.}}: Bare Demo of IEEEtran.cls for Computer Society Journals}
\makeatletter
\def\ps@IEEEtitlepagestyle{%
  \def\@oddfoot{\mycopyrightnotice}%
  \def\@oddhead{\hbox{}\@IEEEheaderstyle\leftmark\hfil\thepage}\relax
  \def\@evenhead{\@IEEEheaderstyle\thepage\hfil\leftmark\hbox{}}\relax
  \def\@evenfoot{}%
}
\def\mycopyrightnotice{%
  \begin{minipage}{\textwidth}
  \centering \scriptsize
  ~\copyright~2024 IEEE. Personal use of this material is permitted. Permission from IEEE must be obtained for all other uses, in any current or future media, including reprinting/ republishing this material for advertising or promotional purposes, creating new collective works, for resale or redistribution to servers or lists, or reuse of any copyrighted component of this work in other works.  A. Alharbi, H. Dong, X. Yi, P. Abeysekara, "Cloned Identity Detection in Social Sensor-Clouds based on Incomplete Profiles" in IEEE Transactions on Services Computing. DOI: 10.1109/TSC.2024.3479912
  \end{minipage}
}
\makeatother

%



\IEEEtitleabstractindextext{%
\begin{abstract}
We propose a novel approach 
to effectively detect cloned identities of social-sensor cloud service providers (i.e. social media users) in the face of incomplete non-privacy-sensitive profile data. Named ICD-IPD, the 
proposed approach first 
extracts account pairs with 
similar usernames or screen names from a given set of 
user accounts collected from a social media.
It then learns 
a multi-view representation associated with a given account and extracts two categories of features for every single account. 
These two categories of features include profile and Weighted Generalised Canonical
Correlation Analysis (WGCCA)-based features that may potentially contain missing values. 
To counter the impact of such missing values,  a missing value imputer will next impute the missing values of the aforementioned profile and WGCCA-based features. After that, the proposed approach further extracts two categories of augmented features for each account pair identified previously, namely, 1) similarity and 2) differences-based features. Finally, these features are concatenated and fed into a Light Gradient Boosting Machine classifier  to detect identity cloning. We evaluated and compared the proposed approach against the existing state-of-the-art identity cloning approaches and other machine or deep learning models atop a real-world dataset. The experimental results show that the proposed approach outperforms the state-of-the-art approaches and models in terms of Precision, Recall and F1-score.
\end{abstract}

}

\maketitle

\IEEEdisplaynontitleabstractindextext

%
\IEEEpeerreviewmaketitle

\IEEEraisesectionheading{\section{Introduction}\label{sec:introduction}}

{\IEEEPARstart{S}{\textit{ocial-sensor cloud services}} (SocSen services) refer to services whose functional (e.g. time and location) and non-functional (e.g. quality and trust) characteristics are abstracted from data (e.g. texts, images, videos, etc.) posted in social media \cite{tooba2022}. These SocSen services can power numerous socially significant and influential applications such as scene reconstruction from social media images, etc. }
 {
The identities of \textit{SocSen service providers} (i.e., individuals that post social media data from social media) have increasingly become a target of the cybercriminals in the recent past 
\cite{gupta2013faking,mendoza2010twitter}.
One such example of these crimes associated with SocSen service provider identities (i.e. social media users) is \textit{identity cloning}, which is an attempt by an adversary to steal the identity information of SocSen service providers to register a fake profile. Many recent 
attempts for identity cloning 
in social media platforms aimed to exploit SocSen service provider 
identities via cloning for either theft for financial fraud or deceiving the public. 
Recent examples illustrate the severity of this problem: Facebook CEO Mark Zuckerberg's account was cloned for financial theft\footnote{https://www.nytimes.com/2018/04/25/technology/fake-mark-zuckerberg-facebook.html}, and a fake Twitter account impersonating Russian President Vladimir Putin gained over one million followers\footnote{https://www.abc.net.au/news/2018-11-29/twitter-suspends-account-impersonating-vladimir-putin/10569064}. These incidents highlight the critical need for effective measures to detect and prevent identity cloning and other malicious activities. Ensuring the security of social media platforms is essential not only for protecting individual identities but also for maintaining the integrity and trustworthiness of online interactions. 
Therefore, it is imperative to put in place measures to detect such attempts to keep attackers at bay and make social media a more secure place for social media users.
Despite its importance, most social media platforms do not offer automated and integrated identity cloning detection. For instance, Instagram and Twitter 
currently \textit{selectively} evaluate identity cloning claims 
only upon receiving legitimate complaints from end-users \footnote{https://help.instagram.com/446663175382270}\footnote{https://help.twitter.com/en/rules-and-policies/twitter-impersonation-policy}. 
However, given the rate at which identity cloning attacks occur, such \textit{selective} approaches can be deemed inadequate to keep social media a safer environment for social media users. Therefore, it is vital to research more \textit{proactive} and \textit{automated} approaches that can also withstand the scale at which social media platforms operate.

Most existing identity cloning detection approaches (such as \cite{devmane2014detection,goga2015doppelganger,kamhoua2017preventing,kontaxis2011detecting}) rely on \textit{complete SocSen service provider (i.e. social media user) profile data}. The performance of these approaches often depends on the availability of comprehensive social media profile information. However, obtaining a comprehensive representation of such profile data is often infeasible due to various reasons. One of the major reasons is that SocSen clouds enable stronger privacy preservation measures not to disclose such information to third-party applications. For example, there has been a growing trend that more \textit{third-party websites/apps} employ \textit{mainstream SocSen cloud APIs} for authentication. These websites/apps can only access limited  profile information authorized by SocSen clouds.
This information is termed as \textit{non-privacy-sensitive} profile information \cite{alharbi1234}.
Our previous research \cite{alharbi1234,alharbi2021privacy} focuses on developing identity cloning detection approaches based on SocSen service providers' \textit{non-privacy-sensitive} profile information. However, SocSen service providers can even opt not to disclose part of the \textit{non-privacy-sensitive} profile information. For example, during account registration, SocSen clouds such as Twitter 
    have made it mandatory that users provide a username, screen name, email address and phone number, which are known as required fields\footnote{https://help.twitter.com/en/using-twitter/create-twitter-account}. The users 
can still \textit{opt out of providing the other optional details}, such as description, location, etc., which can be accessed by Twitter API.

Under such circumstances, cloned user accounts might not expose their full  profile information or \textit{non-privacy-sensitive} profile information in order to reduce the risk of being detected. For example, an adversary can register a cloned profile without including a profile description or adding any post. Therefore, existing identity cloning detection approaches may either fail or perform less in the face of incomplete user profile data since most of the existing approaches are built based on the prerequisite of the existence of the complete profile information or  \textit{non-privacy-sensitive} profile information. According to our experiment results (see Table \ref{perf_cloned}), all the existing identity cloning detection approaches are affected by incomplete profile information. All the existing approaches performed worse when there was incomplete profile information (missing value). 
Imputation is a technique used to handle missing or incomplete data by filling in the gaps with substitute values. Imputation can be performed using statistical or machine learning methods \cite{emmanuel2021survey}. To address these issues, we use imputation methods to replace missing values with appropriate estimates. By applying this technique, we can improve the quality of the data and enhance the detection effectiveness.

To address the above limitations, we propose a novel approach for SocSen service provider \underline{I}dentity \underline{C}loning \underline{D}etection in the face of \underline{I}ncomplete \underline{P}rofile \underline{D}ata (ICD-IPD). ICD-IPD is specially designed to detect cloned identities based on incomplete \textit{non-privacy-sensitive} profile information. ICD-IPD consists of five main components, namely, 1) 
account pair generator (APG), 2) a multi-view learner, 3) a missing value imputer, 4) an account pair feature generator and 5) a prediction model. From a given set of social media users, the \textit{APG} generates account pairs that share similar screen names or usernames. The \textit{multi-view learner} then combines multi-view information of an account to improve learning performance. More specifically, it extracts 
profile (i.e. friends and posts count etc.) and Weighted Generalised Canonical Correlation Analysis (WGCCA)-based features (i.e. combination of multi-view) from a SocSen service provider's \textit{non-privacy-sensitive} profile information. Next, the missing value imputer imputes the missing feature values 
associated with profile and WGCCA-based features. The account pair feature generator then extracts similarity and differences-based features for each account pair in terms of the imputed feature values. Finally, ICD-IPD utilises a Light Gradient Boosting Machine (LightGBM) model atop 
a concatenated form of the aforementioned features to predict whether a pair of accounts compared possibly consists of a cloned account and a victim account. 
Our main contributions can be summarized as follows:
\begin{itemize}
    \item We propose a novel approach to detect SocSen service providers’ identity cloning based on incomplete \textit{non-privacy-sensitive} profiles. To the best of our knowledge, this is the first work in the field of social media identity deception information that specifically works  on user profiles with missing values (incomplete profile data).
    \item We utilize an imputation approach to impute the missing value of incomplete \textit{non-privacy-sensitive} profile data. The utilised imputation approach can substantially enhance the cloned identity prediction performance as shown in Section \ref{eval}.
    \item We adopt an effective prediction model for detecting cloned identities with missing \textit{non-privacy-sensitive} profile information. 
    The proposed prediction model shows better performance than the state-of-art cloned identity detection approaches 
    as well as several other candidate machine and deep learning models.
    \item We 
    present the results of our extensive experiments 
    carried out atop a real-world dataset. The experimental findings showed that ICD-IPD outperforms current cloned identity detection approaches 
    on the Key Performance Indicators: Precision, Recall, and F1-score.
\end{itemize}

The remainder of the paper is structured as follows. Section \ref{re_work} reviews the related work on identity cloning detection. Section \ref{over_sol} 
elaborates our proposed approach to address the challenges outlined previously.  
Meanwhile, Section \ref{eval} 
provides comprehensive details on the methodology used to evaluate the proposed approach and outcomes. Section \ref{concl} concludes the paper.

\section{Related Work}\label{re_work}
\subsection{Applications of Social-Sensor Cloud Services}
{SocSen services are integral to managing and analysing social media data for a variety of applications. Recent advancements highlight the growing complexity and scope of challenges in this domain. Aamir et al. \cite{aamir2017social,aamir2017social1} developed frameworks for selecting SocSen services, specifically targeting scene-related social media images. Their work emphasizes organizing these images based on functional and non-functional attributes, as well as spatial, temporal, and contextual dimensions. Aamir et al. \cite{9001217,aamir2018social} further explored SocSen services enabled scene analysis by proposing models for service composition. These models reconstruct complex scenes by integrating spatio-temporal, textual, and visual features from social-sensor data. Hinduja et al. \cite{hinduja2022machine} proposed a framework to enhance the capabilities of SocSen services by leveraging social media data for early and proactive mental health monitoring, overcoming the limitations of traditional health surveillance systems}

\subsection{Identity Cloning Detection Techniques}
A survey \cite{alharbi2021social} on social media identity deception reveals that various techniques have been proposed for detecting fraudulent accounts and spammers on social media. 
These techniques 
mostly employed  
behavioural features of users such as writing styles for detecting fraudulent activities \cite{masood2019spammer,zheng2015detecting}. However, in 
the context of identity cloning, one of the goals of an attacker 
is to mimic the behavioural profile features to reduce the risk of being detected. Therefore, the aforementioned approaches are less likely to work in our problem setting. Furthermore, some existing works employed 
the \textit{trustworthiness amongst social media users represented by social-network connections amongst them}. These works 
assume that a spammer/fake user cannot develop an arbitrary number of trusted 
connections with legitimate users \cite{al2017sybil,masood2019spammer}. This assumption might 
not always be valid in the context of identity cloning since an attacker can 
clone legitimate user 
profiles and more easily succeed in gaining the trust of other legitimate users. 

The detection of identity cloning on social media has been examined using a variety of approaches. 
{Vyawahare and Govilkar \cite{vyawahare2022fake} developed a method to detect fake and cloned profiles by extracting key attributes (e.g., username, friend count, gender) and calculating a similarity index. Profiles with high similarity scores above a threshold are flagged as potential clones. Jethava and Rao \cite{jethava2022novel} introduced a defensive approach to protect against identity cloning. Their method uses similarity measures (e.g., attribute and friend list) to differentiate between cloned and legitimate users. The approach is implemented on the social app server, where friendship requests are checked for authenticity before being approved.}
Alharbi et al. \cite{alharbi2021privacy} proposed an identity cloning detection strategy based on a deep forest model. 
The aforementioned work extracted an account pair feature representation and a multi-view account representation. These two representations were, then, combined and fed to a deep forest model to predict if a given pair of accounts has a cloned account. Alharbi et al. \cite{alharbi1234} also  proposed an approach that computes the cosine similarity of a pair of accounts based on a learnt single-embedding to detect identity cloning. The aforementioned single-embedding was formulated by merging different views (i.e. posts, network information and profile attributes) extracted from each social media account compared. Goga et al. \cite{goga2015doppelganger} proposed an approach for detecting impersonation. It determines whether or not two accounts are duplicates. Kontaxis et al. \cite{kontaxis2011detecting} introduced a mechanism by which users can ascertain if they have fallen victim in a cloned identity attack. Jin et al. \cite{jin2011towards} studied the behaviour of attackers for identity cloning. The aforesaid work presented two approaches to identify suspicious profiles based on profile similarity. Furthermore, another identity cloning detection approach that detects cloned identities in both single- and cross-platform settings was developed by Devmane and Rana \cite{devmane2014detection} to search for similar, yet cloned, user accounts. Kamhoua et al. also \cite{kamhoua2017preventing} compared user profiles across social media platforms to prevent cloned identities. These existing works depend on the fundamental assumption that access to complete data profiles of social media users is available for identity cloning detection. Therefore, they may either fail or perform less in the face of incomplete user profile data.

According to a recent survey on social media identity deception, most of the social media identity deception detection techniques 
rely on the complete 
profile data of SocSen service providers 
\cite{alharbi2021social}. For example, cloned, fraudulent and spammer accounts on social media can easily 
leave out 
some of the profile information 
in the social network. Since most of these techniques depend on the complete data profiles of social media users, there is an urgent need 
of alternative approaches to detect 
such malicious accounts. 
In reality, the majority of the users on social media platforms have incomplete user profiles due to various reasons (e.g. privacy concerns)\footnote{https://www.statista.com/statistics/934874/users-have-private-social-media-account-usa/}. Thus, utilising existing approaches to detect cloned identities based on incomplete user profiles cannot be assured to perform well, as such incomplete information potentially violates the aforementioned fundamental assumption thereby rendering these existing solutions obsolete. Therefore, we aim to propose an identity cloning detection approach that performs well even in the midst of incomplete user profile data.

\section{Proposed Approach}\label{over_sol}
This section presents a detailed overview of the proposed ICD-IPD approach and its key components.
\subsection{Overview}
\begin{figure*}[h!]
  \centering
  \includegraphics[width=\textwidth]{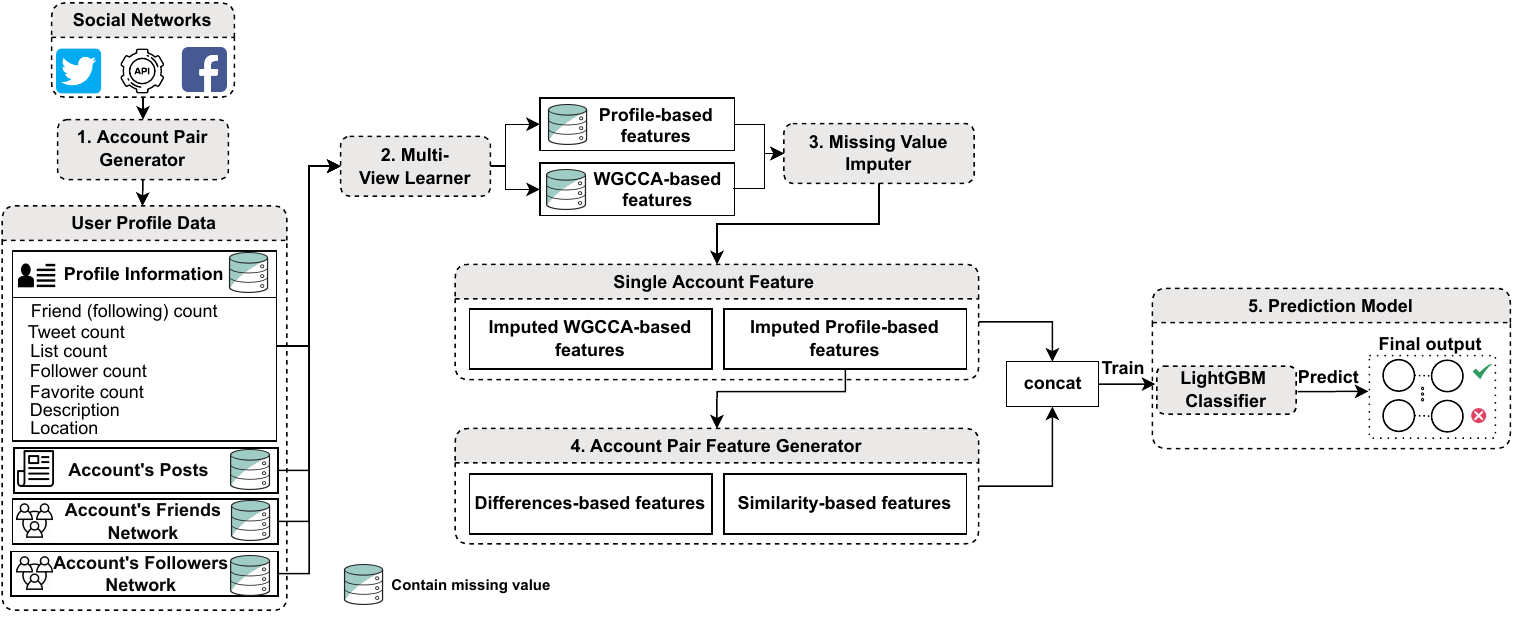}
  \caption{The overview of the ICD-IPD.}
  \label{over}
\end{figure*}

ICD-IPD aims to detect cloned identities with incomplete \textit{non-privacy-sensitive} profile data. As shown in Fig. \ref{over}, ICD-IPD consists of five main components, namely, 1) account pair generator (APG),  2) multi-view learner, 3) missing value imputer, 4) account pair feature generator and 5) prediction model. The \textit{APG} generates 
pairs of accounts with 
similar usernames or screen names from a given set of social media users. Then, the \textit{multi-view learner} aims to 
combine information from several views. 
It extracts two categories of features from the \textit{non-privacy-sensitive} profile information of every single account
, namely, 1) profile and 2) WGCCA-based features. These features can potentially contain 
missing values. Next, to counter the impact of such missing values,  the \textit{missing value imputer} imputes the missing feature values 
of the profile and WGCCA-based features. After imputing the missing values, the account pair feature generator extracts two categories of features for each 
pair of accounts, in the form of  1) similarity and 2) differences-based features. Then, we concatenate the single account feature and account pair feature. Finally, we employ a LightGBM model atop the concatenated features to predict whether a given pair of accounts 
consists of a cloned account and a victim account.

The remainder of this Section is structured as follows. Section \ref{account_pair_g} explains the specifics of the APG. Section \ref{mvl} delivers the details of the multi-view learner while Section \ref{imputer} explains the implementation of the missing value imputer. Section \ref{feature_extraction} 
elaborates the procedure of the account pair feature
generator. Finally,  Section \ref{model} 
introduces the prediction model used.

\subsection{Account Pair Generator (APG)}\label{account_pair_g}
Given a set of social media users, 
APG 
aims to generate 
pairs of accounts where each pair possibly 
contains a cloned account and its victim. An adversary is more likely to register a cloned account sharing the same username or screen name as the victim account \cite{alharbi2021privacy}. Therefore, 
the APG utilizes a method commonly used in prior works, which searches for a pair of users with similar usernames or screen names \cite{alharbi2021privacy,goga2015doppelganger} 
to generate pairs of accounts for identity cloning detection. The aforementioned method uses a similarity score (e.g. cosine similarity) as a metric to decide if two accounts are similar to each other. 
In our implementation,  
the APG pairs up two accounts 
when the similarity score of the associated usernames or screen names of two connected accounts is over 0.8 according to the work of \cite{alharbi1234,alharbi2021privacy}.

\subsection{Multi-View Learner} \label{mvl}
The purpose of the multi-view learner is to improve learning performance by combining multiple views. 
Many data are often gathered through various measurement techniques since a single point of view cannot fully convey the information of all data samples. For example, in social media, users' posts and their networks (e.g. friendship networks) are two distinct types of features that may be considered 
two views derived from the underlying data.
Therefore, we aim to construct a multi-view account representation for a particular user account by merging multiple views that correspond to the account's \textit{non-privacy-sensitive} profile data.  {There are three views that we use for this purpose, namely, 1) profile-based features (providing foundational context about the account), 2) account’s posts (capturing the user's engagement and content creation patterns) and 3) account’s friends and follower networks (providing insights into the user's social interactions and connection).} 
a single embedding is learned from various views using Weighted Generalised Canonical Correlation Analysis (WGCCA). The following sub sections provide further details related to each of the aforementioned views introduced.

\subsubsection{Profile-based Features} \label{profile_info}
We gather 12 profile features (see Table \ref{multi_feature}) to construct a feature vector for each 
individual account within an account pair generated by the APG. These features can be used to describe the user activity and trustworthiness of an account \cite{goga2015doppelganger}. For example, the number of posts/tweets in a social media account can represent a user's level of activity, whereas the number of friends can reflect a user's trust \cite{goga2015doppelganger}. All these features base on the information that can be obtained from mainstream social media APIs, e.g. Twitter API.

\subsubsection{Account's Posts} 
 {For every single account $u$, we use 
Sentence-BERT (SBERT) to retrieve 
a pre-trained language representation \cite{reimers-2019-sentence-bert}. We gathered 
their publicly available posts, denoted as $P = (p_1, ..., p_n)$. We represent each post $p_{i} $($i \in 1,..,n$) using a pre-trained language representation. These pre-trained models are notably effective in extracting text representations relevant to any given task (e.g. categorising etc. \cite{devlin2018bert}). We then tokenize each post $p_{i}$ into 
individual words $w_{i}$. 
A tokenized post is then marked with the [CLS] and [SEP] tags to denote the beginning and end of a phrase. Next, a 
set of tokenized words is 
passed through BERT to embed fixed-sized sentences. Mean aggregation, which outperforms max and CLS aggregation, is used to construct the final post $P$ representations \cite{reimers-2019-sentence-bert}. 
The dimensionality of the representation SBERT outputs for each post is 385, which is BERT’s default output size. We finally compute the mean of all $P$ for every single account $u$.}
\subsubsection{Account's Friends and Follower Networks} We collected information on the friends and followers networks for every 
individual account in an account pair using publicly available information within the underlying social network. We then used Node2Vec \cite{10.1145/2939672.2939754} to learn the corresponding network representation. Node2vec is a well-known approach for unsupervised graph representational learning. It employs a biased random walk approach to maximise the log-probability between two nodes or accounts with an edge between them.
 {Node2vec generates low-dimensional embeddings for users by simulating biased random walks through the user connections. The transition probability from user $v$ to user $x$, given that the previous user was $t$  
(i.e.  the user or node that the random walk visited immediately before the current user $v$) is defined as $\pi v_x = \alpha_{ pq} (t,x) * w_{vx} $, where $w_{vx}$ represents the weight of the connection between users $v$ and $x$, and $\alpha_{ pq} (t,x)$ is a bias factor. The bias factor $\alpha_{ pq} (t,x)$ is $\frac{1}{p}$ if $x$ is the previous user $t$, 1 if $x$ is a direct connection of $t$, and $\frac{1}{q}$ otherwise. Here, $p$ and $q$ are  parameters that control the behavior of the random walk and influence the bias factor in the transition probability calculation.}

\subsubsection{Weighted Generalised Canonical Correlation Analysis (WGCCA)} \label{single_wgcca} The knowledge in posts, friends and follower networks may be utilised to identify cloned accounts \cite{alharbi2021privacy}. Using each representation independently might result in the loss of significant information compared to using a concatenated representation of them. A straightforward and basic strategy is to concatenate all representations together. Such a concatenation strategy, however, might lead to overfitting on smaller training datasets due to the typically higher dimensionality of the account representations.  {Concatenating all representations together increases model complexity and the number of parameters, leading to overfitting on smaller datasets, as the model may capture noise instead of generalizable patterns \cite{ng2004feature}. This phenomenon occurs due to the curse of dimensionality and overparameterization, which result in models that perform well on training data but poorly on testing data \cite{goodfellow2016deep}.} Another reason is that the resultant model could ignore the important information included in each representation since each representation has distinct statistical features. As a result, we use generalised canonical correlation analysis (GCCA), which is a method for learning a single embedding from multiple representations. There are several GCCA variations proposed in the existing literature, such as \cite{carroll1968generalization,robinson1973generalized,tenenhaus2011regularized}. Out of these approaches, Carroll \cite{carroll1968generalization}'s GCCA is a computationally simple and efficient method and thus, we employ that in the proposed approach. Equation \ref{gcca} shows the objective function of the GCCA formulation.

\begin{equation}
\label{gcca}
    arg \min_{G_i,U_i} \sum_i \parallel G-X_iU_i \parallel_F^2 \qquad s.t. G' G = I 
\end{equation}
where  $U_i \in \mathbb{R}^{\ d_i\times k}$ maps from the latent space to the observed feature vector $i$, \(X_i \in \mathbb{R}^{\ n\times d_i}\) represents the  data array 
of the $i^{th}$ feature vector and $G \in \mathbb{R}^{\ n\times k}$ includes all embedded learnt accounts. In the identity cloning detection, each feature vector could have high or less important information. Consequently, we utilise weighted GCCA (wGCCA) that adds weight
$w_i$, which implies the importance of the feature vector, for each feature vector $i$, as shown in Equation \ref{wgcca}.  The columns of $G$ are the eigenvectors of $\sum_i w_i X_i({X_i}'X_i)^{-1}{X_i}'$ and the solution for $U_i=({X_i}'X_i)^{-1}{X_i}'G$.
\begin{equation}
\label{wgcca}
    arg \min_{G_i,U_i} w_i\sum_i \parallel G-X_iU_i \parallel_F^2 \; s.t. G' G = I, w_i\geq0 
\end{equation}

Overall, the multi-view learner extracted a total of 16 features, which are listed in Table \ref{multi_feature}. However, some of these features have missing values. Therefore, in the next component, we elaborate on the strategy used to impute the missing values of profile and WGCCA-based features.

\begin{table*}[h!]
\caption{Single account features and their descriptions, where \halfcirc \  denotes that the feature may contain missing values (i.e. NaN) and \emptycirc \ means the opposite.
}
\label{multi_feature}
\resizebox{\textwidth}{!}{%
\begin{tabular}{p{0.25\textwidth}|l|p{0.18\textwidth}|p{0.7\textwidth}|c}
\toprule
\textbf{Feature category}& No. & \textbf{Features} & \textbf{Description} & \textbf{Missing} \\ \toprule

 \multirow{2}{*}{Profile-based features}&1 & Friend (following) count & The number of accounts that the user follows. & \halfcirc\\ 
&2 & Follower count & The number of users who follow the account. & \halfcirc\\ 

&3 &  Account age & The length of time the account has been open, is expressed in months from the registration date. &  \emptycirc \\
&4 & Tweet count & The number of posts the account has published, including reposts. & \halfcirc\\ 
&5 & List count & The number of lists to which the account is subscribed. & \halfcirc \\ 
&6 & Favorite count & The number of posts that the account has liked. & \halfcirc \\ 
&7 &  Profile URL & A boolean value shows whether or not the account's profile has a URL.  &  \emptycirc \\ 
&8 & Profile image & A boolean value that indicates whether or not the account has submitted a profile picture and instead just uses the default image. &  \emptycirc  \\ 
&9 & Profile background & A boolean value that indicates whether or not the profile background or theme has changed. &  \emptycirc  \\ 
&10 &  Has profile description & A boolean value shows whether or not the account's profile has a description. &  \emptycirc  \\ 
&11 &  Description length & The account's description length. & \halfcirc \\ 
&12 &  Screen name length & The account's screen name length. &  \emptycirc \\ 
  \toprule

\multirow{2}{*}{WGCCA-based features}&13 &  WGCCA$_A$ &  The output of combining the account’s profile, post, friends and follower network. & \halfcirc\\ 
                  &14& WGCCA$_B$  &  The output of combining the account’s profile, post, friends and follower network. & \halfcirc\\
                  &15& WGCCA$_C$  &  The output of combining the account’s profile, post, friends and follower network. & \halfcirc\\
                  &16& WGCCA$_D$  &  The output of combining the account’s profile, post, friends and follower network. & \halfcirc\\                  
                    \toprule                    
                    
\end{tabular}%
}
\end{table*}

\subsection{Missing Value Imputer}\label{imputer}
During account registration, most social media platforms (e.g. Twitter, etc) have made it compulsory that users add a username, screen name, email address and phone number, which are known as required fields. On the other hand, users have been allowed to leave optional fields (e.g. description, location, etc.) empty. An attacker can easily exploit such a setting to avoid being detected via the existing approaches as discussed in detail in the 
Section \ref{sec:introduction}.  

The most popular approach for dealing with missing values in a dataset is missing value imputation. It is the process of replacing a missing value with a suitable substitute value using statistical (e.g. mean) or machine learning (e.g. kNN) approaches \cite{emmanuel2021survey}. The deletion approach, on the other hand, is another approach to 
deal with 
such missing values. However, when the 
percentage of records with missing values in a dataset surpasses 15\%, it is recommended that another approach 
be considered since deleting a data sample missing values might affect the analysis or prediction results \cite{acuna2004treatment}. Therefore, we used missing value imputation approach to deal with the missing values arising in our problem setting.

Our goal is to impute the missing data for the accounts that do not have their complete profile features available. Here, we impute the missing data from the profile and WGCCA-based features introduced in Section \ref{mvl}. 
To that end, we employed Copula-EM \cite{zhao2020missing}, which models data as samples from a Gaussian copula model. This semi-parametric model learns the marginal distribution of each feature value to match the empirical distribution but depicts interactions between feature values using a joint Gaussian distribution. This allows quick inference, confidence interval imputation, and multiple imputations. Copula-EM fits a Gaussian copula model on a dataset with missing values and uses the fitted model to impute the missing value. The Gaussian copula is a modeling approach that uses modifications of latent Gaussian vectors to represent complex multivariate distributions. More specifically, it assumes that the complete data $x \in \mathbb{R} ^p $ is generated as a monotonic transformation of a latent Gaussian vector $z$:
\begin{equation}
\label{a1}
    x = (a_1,..,a_p) = (f_1(z_1),..,f_p(z_p)):=f(z), \text{ for } \, \textbf{z}\sim \mathcal{N}(0,\Sigma) 
\end{equation}
where $x$ is the single account in the account pairs,$a$ is a vectorized form of the 16 features that are explained in Section \ref{mvl} and shown in Table \ref{multi_feature} and $p$ denotes an individual feature in $a$. 

The marginal transformations $ f_1,..,f_p : \mathbb{R} \rightarrow \mathbb{R}$ match the distribution of the observed feature value $x$ to the transformed Gaussian $f(z)$ and are uniquely identifiable given the cumulative distribution function (CDF) of each feature value $x_j$. This model separates the multivariate interaction from the marginal distribution, since the monotone $\textbf{f}$ creates the mapping between the latent variables and the observable variables, whereas $\Sigma$ completely describes the dependent structure. It indicates that $x$ follows the Gaussian copula model with marginal $\textbf{f}$ and copula correlation $\Sigma$ as $x \sim GC(\Sigma,f)$. In other words, Copula-EM constructs a Gaussian copula random vector $x$ by first drawing a latent Gaussian vector $z$ with mean $0$ and covariance $\Sigma$, followed by applying the elementwise monotone function $\textbf{f}$ to $\textbf{z}$ to get $\textbf{x}$. If the CDF for $x_j$ is given by $F_j$, then $f_j$ is uniquely established: $f_j = F^{-1}_j \circ \Phi$, where $\Phi$ denotes the CDF of a standard normal variable. 

Copula-EM models incomplete mixed data (e.g. ordinal, continuous, etc.). Therefore, when a feature value is ordinal, $f_j$ is considered a monotonic step function. Meanwhile, when a feature value is continuous, $f_j$ is strictly monotonic. Copula-EM categorizes a count feature value to one of the above variable types based on its distribution. For the ordinal feature values ($x_j$), CDF $F_j$ and thus $f_j$ is a monotonic step function, and therefore, $f^{-1}_j(x_j) := \{z_i: f_j(z_j) = x_j\}$ is an interval. If $x \sim GC(\Sigma,f)$ is observed at $O$, Copula-EM maps the conditional mean of $\textbf{z}_M$ given observation $X_O$ through $f$ to impute the missing values $\textbf{X}_M$ as follows: 
\begin{align}
\label{gcimpute_eq}
\begin{split}
        \hat{X}_M & =  f_M(\mathbb{E}[z_M \mid X_O,\Sigma ,f]) \\  
    & = f_m(\Sigma_M, O\Sigma^{-1}_O,O\mathbb{E}[z_O\mid X_O,\Sigma,f] )
\end{split}
\end{align}

Copula-EM employs an expectation-maximization (EM) algorithm to estimate the copula correlation matrix $\Sigma$. Given the observed entries $X_O$, the EM algorithm computes the expected covariance matrix of the latent variables $\textbf{z}^i$ at each E-step, as shown in Equation \ref{q123}. The M-step finds the maximum likelihood estimate for the correlation matrix of $\textbf{z}^i$: it updates the model parameter $\Sigma$ as the correlation matrix associated with the expected covariance matrix computed in the E-step.
\begin{equation}
\label{q123}
   \frac{1}{n}\sum_{i=1}^n E[z^i(z^i)^\top \mid X_O]
\end{equation}

\subsection{Account Pair Feature Generator}\label{feature_extraction}
Once we impute the missing profile data, a set of features are extracted for each account pair. The extracted features can be classified into 
two main categories: 1) similarity-based features and 2) difference-based features. We postulate that these two categories of features together can distinguish a cloned account from a genuine 
account more powerfully than when using each one of them individually. We discuss each of these categories of features in-depth in the following subsections.

\subsubsection{Similarity-based features:}
We extract similarity-based features to compare the similarity of the textual features between the account pair such as location, screen name, username, etc. Each feature is given a value within the interval [0,1]. For instance, the screen name similarity score of 1 means the two accounts being compared have a 100\% match on the screen name. 0, on the other hand, signifies that there is no textual resemblance between the two accounts.
We elaborate 
the semantics of computing the aforesaid textual similarity, below.
\paragraph*{\textit{Username, screen name and location similarity}}
Jaro–Winkler string similarity (JS) has been 
shown to perform better on the features 
carrying named value (e.g., property name, username, etc.) \cite{christen2012data,cohen2003comparison}. Therefore, we use JS to compute the similarity of the 
textual features (i.e. location, username, etc.) between the accounts of an account pair, as shown in Equation \ref{qwer}.
\begin{equation}\label{qwer}
JS = \begin{cases}\frac{1}{3}. \frac{m}{\mid S1\mid}+ \frac{m}{\mid S2\mid} +  \frac{m-t}{\mid m \mid} & if :m>0\\0 & :otherwise \end{cases}
\end{equation}
where \(m\), \(t\), \(\mid S1\mid\) and \(\mid S2\mid\) is the number of characters that match, half the transpositions number, and the lengths of the two strings. Matching characters are identical characters in two strings separated by no more than \(w = \frac{max(\mid S1\mid, \mid S2\mid)}{2}\).
JS employs a prefix scale \(p\) that yields a more precise result when two strings share a prefix up to a defined maximum length \(l\).
 \begin{equation}\label{fl11}
         \begin{split}
   JaroWinkler = p+l\times(1- JS) + JS
   \end{split}
 \end{equation}
 
\paragraph*{\textit{Description similarity}}
Users often provide a brief description of themselves in their social media profiles, which typically includes their affiliations with groups, employment, and hobbies. 
This motivated us to compute the similarity on the descriptions of the accounts in a given pair of accounts.
We first transform the textual description to lowercase and remove any punctuation marks and stop words. We then use Term Frequency-Inverse Document Frequency (TF-IDF) to convert the description  
of each account in the account pair into vectors \cite{soucy2005beyond}.
\begin{equation}\label{fl13431}
         \begin{split}
   \cos(\theta) = \frac{\bf USER_A\cdot\bf{USER_B}}{||{\bf USER_A}||\cdot||{\bf USER_B}||}
   \end{split}
 \end{equation}
where \(\bf USER_A\) and \(\bf USER_B\) are the TF-IDF scores of the descriptions for the account pair.

\subsubsection{Differences-based features:}
We extract the differences-based features to compare the public profile features such as the number of posts, followers, etc. that distinguish distinct accounts. 
We assume that the differences between the public profile of the account pair that consists of cloned and victim account will be greater than the differences between any other account pair. For instance, a higher score of difference in the number of posts may suggest the presence of a pair of cloned account and its victim.

Overall, we extracted a total of 10 features from the aforesaid two categories, which are listed in Table \ref{account_pair_feature}.

\begin{table*}[t]
\caption{Account pair features and their descriptions.}
\label{account_pair_feature}
\resizebox{\textwidth}{!}{%
\begin{tabular}{p{0.2\textwidth}|l|p{0.2\textwidth}|p{0.65\textwidth}}
\toprule
\textbf{Feature category}& No. & \textbf{Features} & \textbf{Description} \\ \toprule
 \multirow{2}{*}{Similarity-based features}&1 & Username similarity  & The similarity score of the username for the account pair. \\  
                  &2& Screen name similarity  & The similarity score of the screen name for the account pair.  \\ 
                  &3& Location similarity &  The similarity score of the location for the account pair.   \\ 
                  &4& Description similarity & The similarity score of the description for the account pair.  \\ 
                  &5& Followers Ratio  & The ratio of the number of followers between the account pair. \\ \toprule

\multirow{2}{*}{Differences-based features}&6 &  Followers differences &  The computed score of the difference in the number of followers between the account pair. \\ 
                  &7& Friends differences  & The computed score of the difference in the number of friends between the account pair.   \\
                  &8& Tweets differences   &  The computed score of the difference in the number of tweets between the account pair.  \\ 
                  &9& Favorite differences  & The computed score of the difference in the number of favorite between the account pair.   \\
                  &10& Account age differences  &  The computed score of the difference in the number of account age between the account pair.  \\ 
                    \toprule
\end{tabular}%
}
\end{table*}

\subsection{Prediction Model}\label{model}
We imputed the missing data from the profile and the WGCCA-based features for each individual account, as described in Section \ref{imputer}. We also extracted the similarity and differences-based features for each account pair, as described in Section \ref{feature_extraction}. Next, we concatenated the imputed profile and WGCCA-based features and extracted similarity and differences-based features. We then employ Light Gradient Boosting Machine (LightGBM) \cite{ke2017lightgbm} to predict whether a pair of accounts consists of a cloned and its victim account. LightGBM is a new framework based on a Gradient Boosting Decision Trees (GBDT) \cite{ke2017lightgbm}. GBDT is an ensemble algorithm 
of which the base classifier is a decision
tree. The objective of each iteration of the decision tree is to minimise  
a loss function.

Let $\mathcal{D}=\left\{x_i,y_i\right\}^n_{i=1}$ denote a set of $n$ account pairs, with $\mathcal{X}=\left\{x_i\right\}^n_{i=1}$ denoting the accounts pair representation and $\mathcal{Y}=\left\{y_i\right\}^n_{i=1}\subset \left\{0,1\right\}^n$ denoting the corresponding labels indicating whether or not the account pair contains a cloned account and its victim.  The decision tree model divides each node based on the most informative characteristic (with the largest information gain). The information gained with GBDT is often assessed by the variance after splitting. The variance gain of splitting feature $j$ at point $d$ for a node is defined in Equation \ref{a1111}. 

\begin{align}
\label{a1111}
V_{j\mid O}(d)  = \frac{1}{n_O} & \left( \frac{(\sum_{\left\{x_i \in O:x_{i \ j \ \leq} d\right\}}g_i)^2 }{n{_l}^j_{\mid O}(d)}  \right. \nonumber\\
&\left. \quad + \frac{(\sum_{\left\{x_i \in O:x_{i \ j \ >} d\right\}}g_i)^2}{n{_l}^j_{\mid r}(d)} \right)
\end{align}

\begin{equation}
    n_O = \sum \ I[x_i \in O]
\end{equation}

\begin{equation}
    n{_l}^j_{\mid O}(d) =  \sum \ I[x_i \in O: x_{j  j }\leq d ]
\end{equation}

\begin{equation}
    n{_l}^j_{\mid r}(d) = \sum \ I[x_i \in O: x_{j j }> d ]
\end{equation}
where $O$ denotes the training samples of a decision tree leaf, $n_O$ denotes the number of train samples for a decision tree leaf, $n{_l}^j_{\mid O}(d)$ denotes the number of samples in the decision tree  
of which the initial feature value is less than or equal to $d$, and $n{_l}^j_{\mid r}(d)$ denotes the number of samples in the decision tree with a value larger than $d$ for the second feature.  For feature $j$, the decision tree calculates the $d^*_j = argmax_dV_j(d)$ in order to select the largest information gain $(V_j(d^*_j))$. Then, the data is split into right and left child nodes based on the feature $j^*$.

LightGBM utilises Gradient-based One-Side Sampling (GOSS) to determine the split point via calculating the variance gain. Firstly, we 
sort the gradients of the imputed account 
pair representation based 
on their absolute values in descending order. Secondly, the top$-a \times 100\%$ data samples with larger gradients are selected and kept as a data subset $A$. Then, the remaining imputed account pair samples $A^c$ are randomly sampled to generate another data subset $B$ with size $b \times \mid A^c \mid$. Finally, the instances are split based on the estimated variance gain $\hat{V}_j(d)$ over the subset $A \cup B$ by Equation \ref{a2}.

\begin{align}
\label{a2}
\hat{V}_j(d)  = \frac{1}{n} & \left(  \frac{(\sum_{x_i \in A_l}g_i + \frac{1-a}{b} \sum_{x_i\in B_l}g_i)^2}{n{_l}^j(d)}   \right. \nonumber\\
  &\left. \quad + \frac{(\sum_{x_i \in A_r}g_i + \frac{1-a}{b}\sum_{x_i\in B_r}g_i)^2}{n{_r}^j(d)} \right)
\end{align}
where $A_l = \left\{ x_i \in A : x_{ij} \leq d\right\}$, $A_r = \left\{ x_i \in A : x_{ij} < d\right\}$, $B_l = \left\{ x_i \in B : x_{ij} \leq d\right\}$, $B_r = \left\{ x_i \in B : x_{ij} > d\right\}$, the coefficient ($\frac{1-a}{b}$) is applied to normalize the total of the gradients across $B$ to its original size of $A^c$, and $g_i$ is the negative gradients of the loss function.

Typically, high-dimensional features are mostly sparse and many sparse features are exclusive \cite{ke2017lightgbm}. The sparsity of the feature space allows for minimising the number of features that is almost not useful. Therefore, LightGBM employs an exclusive feature 
bundling approach that can bundle exclusive features that rarely occur simultaneously  
into a single feature. LightGBM generates identical feature histograms for feature bundles and individual features. 

\subsection{Computational Analysis}\label{comp_anal}
 {
\textbf{LightGBM Model:} The initial complexity of the LightGBM algorithm is $O(\#data \times \#features)$, where  $\#data$ is the number of data points and $\#features$ is the number of features. This complexity arises from the need to process each data point for every feature to construct histograms. However, LightGBM uses histogram-based techniques to reduce this complexity. The effective complexity is $O(\#data \times \#bundle)$, where $\#bundle$ is the number of bins or bundles into which feature values are grouped. Since $\#bundle$ is typically much smaller than $\#features$, this approach results in faster training times and improved scalability \cite{ke2017lightgbm}.}

 {
\textbf{Copula-EM Model:} The Copula-EM model, used for imputing missing values, has a time complexity of $O(T \alpha n p^3)$, where $T$ is the number of EM steps required for convergence, $n$ is the number of data points, $p$ is the number of features, and $\alpha$ represents the complexity of the copula function \cite{zhao2020missing}.}

\section{Evaluation}\label{eval}
We conducted a set of experiments to verify the effectiveness of the proposed approach.  The experiments 
were designed to answer the following four key questions:
\begin{itemize}
    \item \textbf{RQ1:} \textit{What imputation approaches are most suitable to counter the effect of incomplete profiles in the proposed approach?} 
    
    We performed a set of experiments to find the best imputation approaches outlined in Section \ref{imp_app} that are most suitable to counter the effect of incomplete profiles, as discussed in Section \ref{rq1}.
    
    \item \textbf{RQ2:} \textit{What are the optimal hyperparameter values of the proposed approach?}
        
    We carried out a set of experiments to assess the impact of the hyperparameters on the proposed approach. The optimal hyperparameter values are depicted in Section \ref{hyper}. We first 
    assessed how the weight $w$ in the wGCCA impacted the employed WGCCA-based features (i.e. post, friends, follower and profile). We then experimented with different weights (i.e. 0.25, 0.5, and 1) for each feature. Each view was given a weight \textit{$[$post, friend, follower, profile$]$}, as discussed in Section \ref{rq2}.

    Next, we performed multiple rounds of experiments to study the impact of the LightGBM model parameters (i.e. \textit{learning\_rate}, \textit{max\_depth} and \textit{num\_leaves}). To 
    that end, we first tested different combinations of \textit{learning\_rate} (i.e. 0.001, 0.01 and 0.1), as discussed in Section \ref{rq2}.
    
    \item \textbf{RQ3:} \textit{How does the proposed approach fare against the state-of-the-art cloned identity detection approaches and other potential candidate solutions?} 

    We carried out a set of experiments to compare the performance of the proposed approach against the state-of-the-art identity cloning detection approaches outlined in Section \ref{cloning_app}. We also evaluated the proposed approach against the machine and deep learning models outlined in Section \ref{ml_app} to justify the use of LightGBM as the predictor, as discussed in Section \ref{rq3}.
    
    \item \textbf{RQ4:} \textit{How impactful is each selected feature on the performance of the proposed approach?} 
    
    We 
    carried out an ablation study of the proposed approach to show the effectiveness of each feature of the proposed ICD-IPD on the overall performance by removing one feature at a time. We compared ICD-IPD with four different variants: 1) ICD-IPD based on the similarity-based features (ICD-IPD$_{SIM}$), 2) ICD-IPD based on differences-based features (ICD-IPD$_{DIF}$), 3) ICD-IPD based on profile-based features (ICD-IPD$_{PROFILE}$) and 4) ICD-IPD based on WGCCA-based features (ICD-IPD$_{WGCCA}$), as discussed in Section \ref{rq4}.
\end{itemize}

\subsection{Experimental Environment}
All the experiments were conducted on a computer with
Intel Core i5 1.80 GHz CPU and 16 GB RAM. All the candidate models compared were implemented in Python. We extracted the pre-trained language representations utilised in ICD-IPD using the SBERT package\footnote{https://github.com/UKPLab/sentence-transformers}. Additionally, we extracted the Node2Vec representations employed by  ICD-IPD using the StellarGraph package\footnote{https://github.com/stellargraph/stellargraph}. We implemented the DL models evaluated using the Python-based Tensorflow\footnote{https://www.tensorflow.org/} library and the other machine learning models evaluated using scikit-learn\footnote{https://scikit-learn.org/stable/}. All experiments were run for 10 rounds with different random permutations of the data. The results were presented as an average computed across all rounds of experiments. 
\subsection{Dataset}
To our knowledge, Twitter is the only prominent social media network that has made public a set of cloned accounts\footnote{https://impersonation.mpi-sws.org/} identified in their platform. This dataset includes 7,015 accounts that have possibly been cloned, and their corresponding victims. Although limited in scope, the existing research evaluated proposed approaches using simulated data.
Similarly, we developed a dataset using the aforementioned Twitter accounts  
in order to evaluate the proposed approach. We collected the user information (i.e profile features, posts, and follower and friends network) via the Twitter APIs\footnote{https://developer.twitter.com/en/docs}. Moreover, most social media platforms state that fake accounts (including cloned accounts) are minority. For example, Twitter estimates fake 
 and spam accounts comprise less than 5\% of users\footnote{https://www.reuters.com/technology/twitter-estimates-spam-fake-accounts-represent-less-than-5-users-filing-2022-05-02/}. Therefore, to mimic a real-world social media environment, where cloned profiles are a minority, we randomly collected 500,000 public Twitter user accounts. Eventually, we developed a dataset that included a total of 514,030 public Twitter profiles for the aforementioned evaluation scenarios.  {Our dataset consists of non-privacy-sensitive profile information, such as publicly available user profiles, posts, and network connections. However, we understand that even public data can pose privacy risks if not managed properly. To address this, we 1) use only data that is publicly shared on Twitter, adhering to ethical standards and not collecting private or sensitive information without consent, and 2) anonymise the data to remove the identifiable details (e.g. names), further protecting user privacy.}

\subsection{Dataset Prepossessing}
The dataset originally contained missing profile data. For example, some of the users do not have any posts or friendship networks. Therefore, since ICD-IPD focuses on detecting cloned accounts with missing profile data, we dropped all accounts that do not have 
complete profile data since we do not have their ground truth for evaluation. We then randomly replaced 50\% of the optional profile features (i.e. description, posts, friends network, etc.) for various percentages of the accounts to simulate missing profile features.  {Here we selected the random replacement percentages of 40\%, 50\%, and 60\% to align with common experimental setups in missing value imputation studies, as demonstrated in \cite{zhao2020missing}.}
In particular, we replaced the values of 10 features of the single account features with NaN. 
We replaced friends count, follower count, favourite count, tweet count, list count and description length from the profile-based features while WGCCA$_A$, WGCCA$_B$, WGCCA$_C$ and WGCCA$_D$ from WGCCA-based features. 
We represent each feature that has missing profile data with (\halfcirc) as shown in Table \ref{multi_feature}. 
Furthermore,  we also used a train-to-test split ratio of 80\%-20\% to train and test the LightGBM 
and other machine learning based predictive models.

\subsection{Evaluation Metrics}
We used Mean Absolute Error (MAE) and Root-Mean-Square Error (RMSE) as the \textit{performance metrics of the imputation approaches}. MAE measures the average magnitude of the errors in a set of predictions without considering their direction. RMSE is a quadratic scoring rule that also measures the average magnitude of the error. Lower scores mean 
the imputation approach 
performs better in the considered experimental setting.
\begin{equation}
    MAE = \sum_{i=1}^{n}|x_i-y_i|
\end{equation}
\begin{equation}
    RMSE = \sqrt{\Sigma_{i=1}^{n}{\frac{(y_i -x_i)^2}{n}}}
\end{equation}
where $n$ is the total number of 
samples, $x_i$ is the true missing value, and $y_i$ is the predicted missing value.

We also used Precision (P), Recall (R) and F1-Score (F1) as the \textit{performance metrics of the cloned account detection approaches}, machine and deep learning models, as well. In the context of the proposed work, Precision is interpreted as the ratio of accurately predicted account pairs (i.e. a cloned account and its related target), while Recall is the ratio of true account pairs that are accurately detected. F1-score is calculated as the harmonic mean of Precision and Recall.

\begin{equation}
    Precision (P) = \frac{Accurately \ Predicted \ Account \  Pairs}{All \ Predicted \ Account \  Pairs}
\end{equation}

\begin{equation}
    Recall (R) = \frac{Accurately \ Predicted \ Account \  Pairs}{Accurate \ Account \ Pairs}
\end{equation}

\begin{equation}
    F1-Score (F1) = 2 \times \frac{Precision*Recall}{Precision+Recall}
\end{equation}

\subsection{Other Approaches Evaluated}
This section describes the different missing value imputation approaches we evaluated to understand how our preferred imputation strategy fares against 
them, as well as how our overall model performs against the other state-of-the-art identity cloning 
and predictive approaches.

\subsubsection{Imputation Approaches:}\label{imp_app}
Here we provide a brief overview of the existing missing value imputation approaches we compared.
\paragraph*{\textbf{Mean}} is a technique that replaces the missing value of a variable with the mean of the available observations.

\paragraph*{\textbf{K-nearest neighbor (KNN)}} is a technique that finds the closest samples in the training set and averages them to fill in 
\textbf{a given} missing value.

\paragraph*{\textbf{MissForest \cite{stekhoven2012missforest}}} is a random forest imputation algorithm that fits a random forest on the observed 
component and predicts the missing 
component.

\paragraph*{\textbf{Copula-EM \cite{zhao2020missing}}} is a technique that fits a Gaussian copula model to impute missing values.

\subsubsection{Existing Identity Cloning Detection Approaches:} \label{cloning_app}
To show the effectiveness of the proposed approach, we compared it 
with the existing state-of-the-art approaches for detecting identity cloning. As part of it,  we used the following existing approaches as baselines:

\paragraph*{\textbf{Basic Profile Similarity (BPS) \cite{jin2011towards}}} This approach examines how much a specific user account and its presumed cloned account overlap in terms of public features and similar friends.

\paragraph*{\textbf{Devmane and Rana \cite{devmane2014detection}}} This approach extracts user accounts' names, workplaces, images, locations, birthdays, education, gender, and friend counts. It then compares these extracted features against a set of user accounts.

\paragraph*{\textbf{Goga et al. \cite{goga2015doppelganger}}} This technique extracts different \textbf{types of} user account features 
\textbf{such as} public features, overlapped 
friends, 
overlap of the time of the tweets (e.g. the difference between the latest tweets) and differences between accounts. A linear kernel is then used to train an SVM classifier, which is subsequently used to identify whether a given account has been impersonated.

\paragraph*{\textbf{Kamhoua et al. \cite{kamhoua2017preventing}}} This technique evaluates the similarity of friend lists and calculates the similarity of features operating \textbf{on} an adjusted similarity measure called Fuzzy-Sim. It examines the following features: name, city, friend list, place of employment age, gender and education. We utilised the same Fuzzy-Sim threshold values (i.e. 0.565 and 0.575) as indicated in the original paper.

\paragraph*{\textbf{Vyawahare and Govilkar \cite{vyawahare2022fake}}} { This approach extracts key attributes from user profiles (e.g., username, friend count, gender) and calculates a similarity index. A Logistic Regression model is then used to determine if an account pair is cloned.}

\paragraph*{\textbf{NPS-AntiClone \cite{alharbi1234}}} This approach extracts different views (i.e. post, network and profile attribute) for each account being compared, and then combines 
the extracted views. Next,  it 
calculates the cosine similarity of the account pair. 
Finally, if the resemblance between a pair of accounts compared is more than 0.1, the account pair is deemed to consist of a  
cloned account and its related target account.

\paragraph*{\textbf{Alharbi et al. \cite{alharbi2021privacy}}} This approach uses similar features to our proposed approach. It then trains the DeepForest model as its predictive strategy.

\subsubsection{Machine and Deep Learning Approaches:}\label{ml_app} We evaluated the LightGBM model against the machine learning and deep learning models mentioned below to justify its use as the cloned identity detection classifier. In the field of cloned identity detection in social media, the following models have been widely used in the literature \cite{alharbi2021social}. Among these models are Random Forest (RF), Adaboost (ADA), Deep Neural Network (DNN), K-nearest Neighbors (KNN), Convolutional Neural Network (CNN), Multi-layer Perceptron (MLP), and eXtreme Gradient Boosting (XGBoost).

Moreover, we also evaluated the LightGBM model based on zero imputation (LightGBM$_{Zero})$. This model imputes zeroes for the missing values. 


\subsection{Hyperparameter Tuning} \label{hyper}
\begin{table}[t]
\centering
\caption{Values of hyperparameter utilised for the candidate machine learning and DL models}
\label{parm}
\resizebox{\columnwidth}{!}{%
\begin{tabular}{p{0.1\columnwidth}p{0.7\columnwidth}}
\toprule
\textbf{Model} & \textbf{Parameter} \\ \toprule
ADA & estimators = 100 \\
RF & estimators = 50 \\
MLP & activation = relu, solver = adam \\
CNN & 8 layers, filters = 64 and 8, kernel size = 2 and 1, pool size = 2 \\
DNN & 6 layers (300, 250,150,100, 50, 1)\\
KNN & neighbors = 15 \\ 
\toprule
\end{tabular}%
}
\end{table}
\begin{table}[t]
\centering
\caption{Values of hyperparameter utilised for the LightGBM model}
\label{lightgbm_parm}
\resizebox{\columnwidth}{!}{%
\begin{tabular}{p{0.15\columnwidth}p{0.20\columnwidth}p{0.10\columnwidth}p{0.10\columnwidth}p{0.10\columnwidth}}
\toprule
\textbf{Model} & \textbf{Parameter} & \textbf{40\%} & \textbf{50\%} & \textbf{60\%} \\ \toprule
LightGBM & \textit{learning\_rate} & 0.1 & 0.1 & 0.1 \\
 & \textit{max\_depth} & 15 & 18 & 10\\
 & \textit{num\_leaves} & 40 & 80 & 80 \\
 & \textit{reg\_alpha} & 0.01 & 0.01 & 0.03 \\
\toprule
\end{tabular}%
}
\end{table}

We followed the existing works \cite{alharbi1234,alharbi2021privacy} and selected an account 
pair when the similarity score of the screen names or usernames of the two is over a 0.8 for the APG. Following the existing works \cite{alharbi1234,alharbi2021privacy}, we also utilised  `all-MiniLM-L6-v2'\footnote{https://huggingface.co/sentence-transformers/all-MiniLM-L6-v2} as the pre-trained model for SBERT, and the dimensions of SBERT and Node2Vec 
were set as $385$ and $128$, respectively. We set the wGCCA's weights $w$ to $[0.25, 1.0, 1.0, 0.25]$, $[1.0, 1.0, 0.5, 0.25]$ and $[0.25, 0.5, 0.5, 0.25]$ for the the account percentages (i.e.
40\%, 50\%, 60\%), respectively. We also set the other parameters of the machine and deep learning models following the existing works (see Table \ref{parm}) \cite{alharbi1234,alharbi2021privacy}. Copula-EM does not require any hyperparameter tuning.

Furthermore, we fine-tuned all the hyperparameters of the LightGBM model to obtain optimal performance. As part of it, we 
experimented with a range of different values of the \textit{learning\_rate} $[0.1, 0.01, 0.001]$.
We also tested different numbers of boosting stages by increasing the \textit{max\_depth} parameter 
within the range $[5, 10, 15, 18, 20]$. Additionally, We experimented with a range of different values of the \textit{num\_leaves} $[40, 60, 80]$. Table \ref{lightgbm_parm} details the values of the hyperparameter utilised to configure the LightGBM based on various percentages (i.e.
40\%, 50\%, 60\%) of the accounts in the underlying dataset.

\subsection{Results and Discussion}
This section presents and discusses the  
results of our experiments described previously. We first discuss the 
results of the experiments concerning  
imputation approaches. We then report 
our findings related to hyperparameter tuning of the proposed approach. Next, we discuss the performance of cloned account detection with respect to the other state-of-the-art approaches as well as machine and deep learning approaches evaluated. Finally, we report the results of an ablation study of the proposed approach conducted in order to measure the impact of each selected individual feature on the overall performance.
\subsubsection{Performance of Imputation Approaches (RQ1)}\label{rq1}

\begin{table*}[h!] 
\caption{Performance of Imputation Approaches} 
\label{impu} 
\setlength\tabcolsep{0pt} 
\begin{tabular*}{\textwidth}{l @{\extracolsep{\fill}}
                            *{6}{S[table-format=1.3]}} 
\toprule
& \multicolumn{2}{c}{\textbf{40\%}} & \multicolumn{2}{c}{\textbf{50\%}} & \multicolumn{2}{c}{\textbf{60\%}} \\
\cmidrule{2-3} \cmidrule{4-5} \cmidrule{6-7} 
\textbf{Approach} & {\textbf{MAE}} & {\textbf{RMSE}}  & {\textbf{MAE}} & {\textbf{RMSE}}  & {\textbf{MAE}} & {\textbf{RMSE}}  \\ 
\midrule
Mean & 0.488 & 1.014 & 0.489 & 1.003 & 0.488 & 0.990	 \\ 
KNN & 0.472 & 1.011 & 0.485 & 1.042 & 0.494 & 1.013  \\ 
MissForest \cite{stekhoven2012missforest} & 0.604 & 1.230 & 0.598 & 1.235 & 0.556 & 1.193\\ 
Copula-EM \cite{zhao2020missing} & \underline{\textbf{0.417}} & \underline{\textbf{0.975}} & \underline{\textbf{0.414}} & \underline{\textbf{0.969}} & \underline{\textbf{0.420}} & \underline{\textbf{0.971}}  \\ 
\bottomrule
\end{tabular*} 
\end{table*}
We evaluated and compared the performance of the missing value imputation approaches to justify the usage of Copula-EM. Table \ref{impu} reports the performance of the missing value imputation approaches based on various percentages (i.e. 40\%, 50\%, 60\%) of the accounts with missing profile features. Copula-EM outperforms all other imputation approaches in both MAE and RMSE for all percentages of the accounts. Copula-EM achieved an MAE of 0.417, 0.414 and 0.420 for 40\%, 50\% and 60\%, respectively. Copula-EM also achieved an RMSE of 0.975, 0.969, 0.971 for 40\%, 50\% and 60\%, respectively. Randomly replacing the optional features of 50\% of the accounts achieved the best-performing results. We attribute the 
superior performance achieved when using Copula-EM to its capability of modeling incomplete mixed data using a Gaussian copula model, as well as, employing an efficient approximation expectation maximization (EM) approach for estimating the copula correlation. In addition, Copula-EM does not require tuning parameters. However, mean imputation does not maintain the correlations between other profile features. In other words, the profile features of an account can be dependent on the missing values themselves.

\subsubsection{Impact of the hyperparameters (RQ2)}\label{rq2}
\paragraph*{\textbf{Impact of the wGCCA's weight $w$}}\label{wgcca_w}
Figure \ref{weight1} displays the top 10 results achieved when applying different weight combinations based on various percentages (i.e. 40\%, 50\%, 60\%) of the accounts with missing profile features. The optimal value of w for each feature is $[0.25,1.0,1.0,0.25]$, $[1.0,1.0,0.5,0.25]$ and $[0.25,0.5,0.5,0.25]$ for 40\%, 50\% and 60\%, respectively. Friend networks have a high impact on the wGCCA for all percentages of the accounts. On the other hand, the posts and profile attributes did not have a high impact on the wGCCA except in 50\% of the accounts for posts which was given 1 as a weight.  

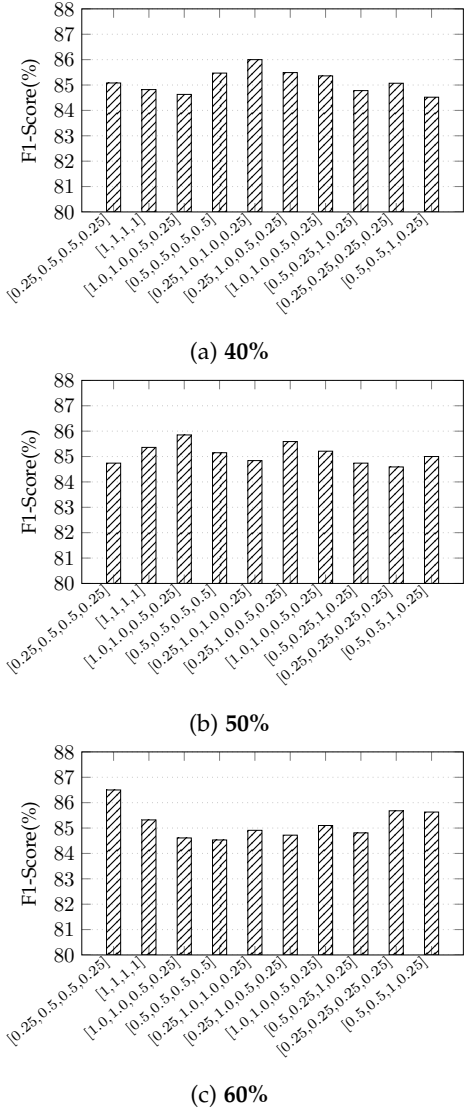
\begin{figure}[t]
\centering
\subfloat[][\textbf{40\%}]{\resizebox{0.35\textwidth}{!}{
  \begin{tikzpicture}[thick,scale=1, every node/.style={scale=1}]
    	\begin{axis}[
        width=8cm,
        height=5cm,
        xtick={0,1,2,3,4,5,6,7,8,9},
        xticklabels={$_{[0.25,0.5,0.5,0.25]}$,$_{[1,1,1,1]}$,$_{[1.0,1.0,0.5,0.25]}$,$_{[0.5,0.5,0.5,0.5]}$,$_{[0.25,1.0,1.0,0.25]}$,$_{[0.25,1.0,0.5,0.25]}$,$_{[1.0,1.0,0.5,0.25]}$, $_{[0.5,0.25,1,0.25]}$, $_{[0.25,0.25,0.25,0.25]}$, $_{[0.5,0.5,1,0.25]}$ },
        x tick label style={rotate=40,anchor=east},
        ytick={80,81,82,83,84,85,86,87,88},
        ymin=80,
        ymax=88,
        ymajorgrids=true,
        grid style=dotted,
        ylabel={F1-Score(\%)},
        every axis plot/.append style={
          ybar,
          bar width=.4,
          bar shift=0pt,
          fill
        }
      ]
      \addplot [pattern=north east lines] coordinates {(0,85.08)};
      \addplot [pattern=north east lines] coordinates {(1,84.82)};
      \addplot [pattern=north east lines] coordinates {(2,84.63)};
      \addplot [pattern=north east lines] coordinates {(3,85.47)};
      \addplot [pattern=north east lines] coordinates {(4,86.00)};
      \addplot [pattern=north east lines] coordinates {(5,85.49)};
      \addplot [pattern=north east lines] coordinates {(6,85.36)};
      \addplot [pattern=north east lines] coordinates {(7,84.78)};
      \addplot [pattern=north east lines] coordinates {(8,85.07)};
      \addplot [pattern=north east lines] coordinates {(9,84.52)};

    \end{axis}
  \end{tikzpicture}
}}

\centering
\subfloat[][\textbf{50\%}]{\resizebox{0.35\textwidth}{!}{
  \begin{tikzpicture}[thick,scale=1, every node/.style={scale=1}]
    \begin{axis}[
        width=8cm,
        height=5cm,
        xtick={0,1,2,3,4,5,6,7,8,9},
        xticklabels={$_{[0.25,0.5,0.5,0.25]}$,$_{[1,1,1,1]}$,$_{[1.0,1.0,0.5,0.25]}$,$_{[0.5,0.5,0.5,0.5]}$,$_{[0.25,1.0,1.0,0.25]}$,$_{[0.25,1.0,0.5,0.25]}$,$_{[1.0,1.0,0.5,0.25]}$, $_{[0.5,0.25,1,0.25]}$, $_{[0.25,0.25,0.25,0.25]}$, $_{[0.5,0.5,1,0.25]}$ },
        x tick label style={rotate=40,anchor=east},
        ytick={80,81,82,83,84,85,86,87,88},
        ymin=80,
        ymax=88,
        ymajorgrids=true,
        grid style=dotted,
        ylabel={F1-Score(\%)},
        every axis plot/.append style={
          ybar,
          bar width=.4,
          bar shift=0pt,
          fill
        }
      ]
      \addplot [pattern=north east lines] coordinates {(0,84.74)};
      \addplot [pattern=north east lines] coordinates {(1,85.36)};
      \addplot [pattern=north east lines] coordinates {(2,85.85)};
      \addplot [pattern=north east lines] coordinates {(3,85.15)};
      \addplot [pattern=north east lines] coordinates {(4,84.84)};
      \addplot [pattern=north east lines] coordinates {(5,85.59)};
      \addplot [pattern=north east lines] coordinates {(6,85.21)};
      \addplot [pattern=north east lines] coordinates {(7,84.74)};
      \addplot [pattern=north east lines] coordinates {(8,84.59)};
      \addplot [pattern=north east lines] coordinates {(9,85.00)};

    \end{axis}
  \end{tikzpicture}
}}

\subfloat[][\textbf{60\%}]{\resizebox{0.35\textwidth}{!}{
  \begin{tikzpicture}[thick,scale=1, every node/.style={scale=1}]
    	\definecolor{coolblack}{rgb}{0.0, 0.18, 0.39}
    \begin{axis}[
        width=8cm,
        height=5cm,
        xtick={0,1,2,3,4,5,6,7,8,9},
        xticklabels={$_{[0.25,0.5,0.5,0.25]}$,$_{[1,1,1,1]}$,$_{[1.0,1.0,0.5,0.25]}$,$_{[0.5,0.5,0.5,0.5]}$,$_{[0.25,1.0,1.0,0.25]}$,$_{[0.25,1.0,0.5,0.25]}$,$_{[1.0,1.0,0.5,0.25]}$, $_{[0.5,0.25,1,0.25]}$, $_{[0.25,0.25,0.25,0.25]}$, $_{[0.5,0.5,1,0.25]}$ },
        x tick label style={rotate=40,anchor=east},
        ytick={80,81,82,83,84,85,86,87,88},
        ymin=80,
        ymax=88,
        ymajorgrids=true,
        grid style=dotted,
        ylabel={F1-Score(\%)},
        every axis plot/.append style={
          ybar,
          bar width=.4,
          bar shift=0pt,
          fill
        }
      ]
      \addplot [pattern=north east lines] coordinates {(0,86.50)};
      \addplot [pattern=north east lines] coordinates {(1,85.32)};
      \addplot [pattern=north east lines] coordinates {(2,84.61)};
      \addplot [pattern=north east lines] coordinates {(3,84.53)};
      \addplot [pattern=north east lines] coordinates {(4,84.91)};
      \addplot [pattern=north east lines] coordinates {(5,84.72)};
      \addplot [pattern=north east lines] coordinates {(6,85.1)};
      \addplot [pattern=north east lines] coordinates {(7,84.81)};
      \addplot [pattern=north east lines] coordinates {(8,85.68)};
      \addplot [pattern=north east lines] coordinates {(9,85.63)};

    \end{axis}
  \end{tikzpicture}
}}
    \caption{Impact of the weight of the wGCCA}
  \label{weight1}
\end{figure}

\paragraph*{\textbf{Impact of the LightGBM model parameters}}
Table \ref{lr} reports the impact of the \textit{learning\_rate} based on various percentages (i.e. 40\%, 50\%, 60\%) of the accounts with missing profile features. The optimal value of the \textit{learning\_rate} is 0.01 for percentages of the accounts.
We also tested different combinations of \textit{max\_depth} (i.e. 5, 10, 15, 18 and 20). Table \ref{depth} reports the impact of the \textit{max\_depth}  based on various percentages (i.e. 40\%, 50\%, 60\%) of the accounts. It can be seen that each percentage of the accounts has a different \textit{max\_depth}. The \textit{max\_depth} is 15, 18 and 10 for 40\%, 50\% and 60\% of the accounts, respectively.
We also evaluated various \textit{num\_leaves} combinations (i.e. 40, 60 and 80). Table \ref{nl} reports the impact of the \textit{num\_leaves} based on various percentages (i.e. 40\%, 50\%, 60\%) of the accounts. It can be observed that 50\% and 60\% have same \textit{num\_leaves} which is 80. For the 40\% of the accounts, the optimal value of the \textit{num\_leaves} is 60.

\begin{table}[]
\caption{Impact of the \textit{learning\_rate} of the LightGBM} 
\label{lr}
\centering
\resizebox{0.9\columnwidth}{!}{%
\begin{tabular}{p{0.15\columnwidth}p{0.08\columnwidth}p{0.10\columnwidth}p{0.10\columnwidth}p{0.10\columnwidth}}
\bottomrule
 \textbf{Parameter} & \textbf{Value} & \multicolumn{3}{c}{\textbf{F1-Score}}                            \\ \bottomrule
 &  & \multicolumn{1}{c}{\textbf{40\%}} & \multicolumn{1}{c}{\textbf{50\%}} & \textbf{60\%}  \\ \cmidrule{3-5}
 
\textit{learning\_rate} & 0.001 & 84.30 & 85.18 & 84.51\\ 
 & 0.01 & 84.88 & 85.44 & 85.34 \\ 
 & 0.1 & \underline{\textbf{86.00}} & \underline{\textbf{85.85}} & \underline{\textbf{86.50}} \\ 
\bottomrule
\end{tabular}
}
\end{table}

\begin{table}[]
\caption{Impact of the \textit{max\_depth} of the LightGBM} 
\label{depth}
\centering
\resizebox{0.9\columnwidth}{!}{%
\begin{tabular}{p{0.15\columnwidth}p{0.08\columnwidth}p{0.10\columnwidth}p{0.10\columnwidth}p{0.10\columnwidth}}
\bottomrule
 \textbf{Parameter} & \textbf{Value} & \multicolumn{3}{c}{\textbf{F1-Score}}                            \\ \bottomrule
 &  & \multicolumn{1}{c}{\textbf{40\%}} & \multicolumn{1}{c}{\textbf{50\%}} & \textbf{60\%}  \\ \cmidrule{3-5}
\textit{max\_depth} & 5 & 84.30 & 85.08 & 84.76\\ 
 & 10 & 84.69 & 84.46 & \underline{\textbf{86.50}} \\ 
 & 15 & \underline{\textbf{86.00}} & 85.35 & 85.54 \\ 
 & 18 & 85.38 & \underline{\textbf{85.85}} & 84.90\\ 
 & 20 & 85.78 & 84.97 & 84.55 \\  \bottomrule
\end{tabular}
}
\end{table}

\begin{table}[]
\caption{Impact of the \textit{num\_leaves} of the LightGBM} 
\label{nl}
\centering
\resizebox{0.9\columnwidth}{!}{%
\begin{tabular}{p{0.15\columnwidth}p{0.08\columnwidth}p{0.10\columnwidth}p{0.10\columnwidth}p{0.10\columnwidth}}
\bottomrule
 \textbf{Parameter} & \textbf{Value} & \multicolumn{3}{c}{\textbf{F1-Score}}                            \\ \bottomrule
 &  & \multicolumn{1}{c}{\textbf{40\%}} & \multicolumn{1}{c}{\textbf{50\%}} & \textbf{60\%}  \\ \cmidrule{3-5}
\textit{num\_leaves} & 40 & \underline{\textbf{86.00}} & 85.08 & 85.45\\ 
 & 60 & 85.70 & \underline{\textbf{85.85}} &  84.16  \\ 
 & 80 & 85.34 & 85.35 & \underline{\textbf{86.50}} \\  \bottomrule
\end{tabular}
}
\end{table}
\subsubsection{Performance of Cloned Account Detection (RQ3)}\label{rq3}
\begin{table*}[ht!] 
\caption{Performance of Cloned Account Detection} 
\label{perf_cloned} 
\setlength\tabcolsep{0pt} 
\begin{tabular*}{\textwidth}{l @{\extracolsep{\fill}}
                            *{12}{S[table-format=1.1]}} 
\toprule
& \multicolumn{3}{c}{\textbf{40\%}} & \multicolumn{3}{c}{\textbf{50\%}} & \multicolumn{3}{c}{\textbf{60\%}} & \multicolumn{3}{c}{\textbf{Complete}}\\
\cmidrule{2-4} \cmidrule{5-7} \cmidrule{8-10} \cmidrule{11-13}
\textbf{Approach} & {\textbf{P}} & {\textbf{R}} & {\textbf{F1}} & {\textbf{P}} & {\textbf{R}} & {\textbf{F1}} & {\textbf{P}} & {\textbf{R}} & {\textbf{F1}} & {\textbf{P}} & {\textbf{R}} & {\textbf{F1}}\\ 
\midrule
BPS \cite{jin2011towards} & 65.44 & 69.80 & 67.55 & 63.51 & 69.68 & 66.45 & 63.74 & 68.42  & 66.00  & 68.31	&75.14&	71.56  \\ 
Devmane and Rana \cite{devmane2014detection} & 64.43 & 68.01 & 66.17 &  62.61 &  70.04 &  66.12 & 68.57 & 70.49  &  69.52 & 64.31 &	77.14 &	70.15 \\ 
Goga et al. \cite{goga2015doppelganger} & 63.42 & 70.36 & 66.71 & 61.19 &  69.04 & 64.87  & 63.74 & 68.42 &  66.00 & 65.85&	73.74&	69.57  \\ 
Kamhoua et al. \cite{kamhoua2017preventing} & 58.56 & 70.72 & 64.07 & 60.05  &  71.80  & 65.40 & 61.64 & 67.79 & 64.57  & 60.17 &	76.66 &	67.42   \\
 {\textbf{Vyawahare and Govilkar \cite{vyawahare2022fake} }} & 72.45  & 77.11  & 74.70  & 73.61   & 74.85   & 74.22  &  71.88 & 74.43 &  73.13 & 71.25 &	77.51  &	74.24    \\
NPS-AntiClone \cite{alharbi1234} & 71.10 & 61.08 & 65.71 & 70.37  &  60.74 & 65.20  & 71.84 & 60.75 &  65.83 & 71.14 & 67.67 & 69.36  \\
Alharbi et al. \cite{alharbi2021privacy} & 93.11 & 78.31 & 85.07 &  92.65  & 76.17 & 83.60 & 92.40 & 77.67 &  84.40  & 91.04 &	74.71 &	82.07 \\
ICD-IPD & \underline{\textbf{94.00}} & \underline{\textbf{79.29}} & \underline{\textbf{86.00}} & \underline{\textbf{93.44}} & \underline{\textbf{79.42}} & \underline{\textbf{85.85}} & \underline{\textbf{94.05}} & \underline{\textbf{80.09}} & \underline{\textbf{86.50}} & \underline{\textbf{94.40}} & \underline{\textbf{79.81}} & \underline{\textbf{86.49}} \\

\bottomrule
\end{tabular*} 
\end{table*}
\begin{table*}[ht!] 
\caption{Performance of Machine and Deep Learning} 
\label{ml_per} 
\setlength\tabcolsep{0pt} 
\begin{tabular*}{\textwidth}{l @{\extracolsep{\fill}}
                            *{12}{S[table-format=1.1]}} 
\toprule
& \multicolumn{3}{c}{\textbf{40\%}} & \multicolumn{3}{c}{\textbf{50\%}} & \multicolumn{3}{c}{\textbf{60\%}} & \multicolumn{3}{c}{\textbf{Complete}}\\
\cmidrule{2-4} \cmidrule{5-7} \cmidrule{8-10} \cmidrule{11-13}
\textbf{Model} & {\textbf{P}} & {\textbf{R}} & {\textbf{F1}} & {\textbf{P}} & {\textbf{R}} & {\textbf{F1}} & {\textbf{P}} & {\textbf{R}} & {\textbf{F1}} & {\textbf{P}} & {\textbf{R}} & {\textbf{F1}}\\ 
\midrule
ADA & 92.26 & 72.61 & 81.26	& 91.29 & 71.14 & 79.96	& 91.79 & 70.39 & 79.67 & 92.19 & 71.69 & 80.66\\ 
CNN & 92.26 & 49.05 & 64.05 & 93.26 & 43.83 & 59.63 & 91.04 & 49.72 & 64.31 & 86.90 & 13.27 & 23.03 \\ 
DNN & 83.06 & 22.88 & 35.88 & 82.81 & 26.22 & 39.83 & 85.61 & 26.11 & 40.02 & 77.58 & 24.61 & 37.36 \\ 
KNN & 67.08 & 21.57 & 32.63 & 66.82 & 20.88 & 31.80 & 68.23 & 20.94 & 32.05 & 77.44 & 26.47 & 39.46 \\ 
MLP & 77.19 & 35.78 & 48.70 & 83.14 & 36.46 & 50.34 & 73.64 & 40.81 & 52.00 &  81.28 & 41.50 & 54.88 \\ 
RF & 94.18 & 73.35 & 82.47 & 93.58 & 72.15 & 81.48 & \underline{\textbf{94.16}} & 70.67 & 80.73 & 94.37 & 73.06 & 82.36 \\ 
XGBoost & \underline{\textbf{94.34}} & 78.44 & 85.66 & \underline{\textbf{94.11}} & 77.72 & 85.13 & 94.11 & 77.70 & 85.03 & \underline{\textbf{94.91}} & 77.61 & 85.39 \\
LightGBM & 94.00 & \underline{\textbf{79.29}} & \underline{\textbf{86.00}} & 93.44 & \underline{\textbf{79.42}} & \underline{\textbf{85.85}} & 94.05 & \underline{\textbf{80.09}} & \underline{\textbf{86.50}} & 94.40 & \underline{\textbf{79.81}} & \underline{\textbf{86.49}} \\ \bottomrule
LightGBM$_{Zero}$ & 93.95 & 77.60 & 84.99 & 93.95 & 77.60 & 84.99 & 94.58 & 77.33 & 85.09 & \textendash & \textendash & \textendash \\ 
\bottomrule
\end{tabular*} 
\end{table*}

Table \ref{perf_cloned} reports the performance of these cloned account detection approaches evaluated. ICD-IPD 
was observed to outperform all the state-of-the-art approaches in terms of Precision, Recall and F1-Score based on all percentages (i.e. 40\%, 50\%, 60\%) of the accounts. More specifically, ICD-IPD achieved a Precision of 94.00\%, 93.44\% and 94.05\% for 40\%, 50\% and 60\% of accounts, respectively. ICD-IPD achieved a Recall of 79.29\%, 97.42\% and 80.09\% for 40\%, 50\% and 60\% of accounts, respectively. ICD-IPD achieved a F1-Score of 86.01\%, 85.85\% and 86.50\% for 40\%, 50\% and 60\% of accounts, respectively. In summary, ICD-IPD with 40\% of the accounts was observed to be 0.89\%, 0.98\% and 0.93\% better in terms of Precision, Recall and F1-score than Alharbi et al. \cite{alharbi2021privacy}, which is the second-best performing state-of-the-art approaches. ICD-IPD with 50\%  of the accounts is also 0.79\%, 3.25\% and 2.25\% better in terms of Precision, Recall and F1-score than Alharbi et al. \cite{alharbi2021privacy}, which is the second-best performing state-of-the-art approaches. ICD-IPD with 60\%  of the accounts is also 1.65\%, 2.42\% and 2.1\% better in terms of Precision, Recall and F1-score than Alharbi et al. \cite{alharbi2021privacy}, which is the second-best performing state-of-the-art approaches. ICD-IPD with 60\%  of the accounts was increased by 0.05\% and 0.61\% in Precision against ICD-IPD with 40\% and 50\% of the accounts, respectively. ICD-IPD with 60\%  of the accounts was observed to be 0.80\% and 0.67\% in Recall against ICD-IPD with 40\% and 50\% of the accounts, respectively. ICD-IPD with 60\%  of the accounts was 0.5\% and 0.67\% higher in F1-score against ICD-IPD with 40\% and 50\% of the accounts, respectively.

We attribute the superior performance of the ICD-IPD to its ability to effectively deal with incomplete data whereas the reliance of the existing detection approaches on the complete profile data could be deemed the reason for their comparatively low performance. For example, BPS \cite{jin2011towards} relies on friends' network similarity and profile information. In this case, the aforementioned approach may perform less when an attacker clones a victim's account without having any friends. Additionally, NPS-AntiClone \cite{alharbi1234} detection approach depends on the accounts 
carrying all profile information including the account's posts and friendship network. Therefore, the reported results show that the proposed approach better fits the scenario 
where the accounts do not have their complete profile data.  {Vyawahare and Govilkar \cite{vyawahare2022fake} detected cloned accounts by assessing the similarity of complete profile data based on a logistic regression model. However, logistic regression can struggle with sparse data, leading to overfitting and challenges in estimating model coefficients \cite{zou2005regularization}.}

\paragraph*{\textbf{Performance of Machine and Deep Learning}}
Table \ref{ml_per} reports the performance of machine and deep learning models based on various percentages (i.e. 40\%, 50\%, 60\%) of the accounts as well as 
degrees of completeness of profile data. It can be seen that the proposed LightGBM model outperformed all of the other candidate machine and deep learning models on Recall and F1-score 
against all percentages as well as 
degrees of completeness of profile data. The proposed LightGBM model based on the 40\% of the accounts were 0.40\%, 0.52\% and 0.48\% lower in Precision, Recall and F1-Score than the LightGBM model based on complete data. Additionally, the proposed LightGBM model based on the 50\% of the accounts were 0.96\%, 0.39\% and 0.64\% lower in Precision, Recall and F1-Score than the LightGBM model based on complete data. 
Interestingly, the proposed LightGBM model based on the 60\% of the accounts was 1.09\% and 0.01\% higher in Recall and F1-Score than the LightGBM model based on complete data. The proposed LightGBM model is specially designed to predict using the imputed data. The complete data might contain noises that can affect the performance results. Additionally,  the Copula-EM estimates the missing value of the features based on the observed features. Thus, it can increase the correlation between the features. On the other hand, the correlation of the features in the complete data is not positive. As shown in Table \ref{ml_per}, the LightGBM model-based on zero imputation 
against all percentages 
was observed to have the lowest performance. 
The reason is that zero imputation can affect the prediction performance of machine and deep learning models \cite{hazan2015classification}. More especially, zero imputation can cause sparsity bias in the training of a predictive model. The predictive model's output varies significantly with regard to the rate of missingness in the provided input, indicating that it has a negative effect on model performance \cite{yi2019not}.

Furthermore, the XGBoost model achieved a higher Precision than the proposed LightGBM 
against all percentages as well as 
degrees of completeness of profile data. However, on the other evaluation metrics, the proposed LightGBM model outperformed all of the other candidate machine and deep learning models. We believe the superiority of the LightGBM model against the other machine and deep learning approaches evaluated stems from the use of Gradient Boosting (GB) in its implementation, which is a technique 
enabling powerful classifiers 
that generally perform very well on structured data. 
Although XGBoost too is a GB-based approach, its use of 
a level-wise tree growth strategy 
can result in many nodes achieving low splitting gains and increasing computations without improving accuracy. On the other hand, the LightGBM adopts a leaf-wise approach, which is both comparatively quicker and more accurate. With the leaf-wise approach, asymmetric and deeper trees are grown by identifying the node with the highest gain at each layer and only splitting that node.

\subsubsection{Ablation Study (RQ4)}\label{rq4}
\begin{figure}[t]
\centering
\subfloat[][\textbf{40\%}]{\resizebox{0.35\textwidth}{!}{
\begin{tikzpicture} 
\begin{axis}[
    ybar,
    legend style={at={(0.5,-0.35)},anchor=north,legend columns=-1},
    width=11cm,
    height=6cm,
    ymin=20,
    ymax=90,
    ytick={20,30,40,50,60,70,80,90},
    enlargelimits=0.15,
    grid style=dotted,
    ylabel={\textbf{Performance} (\%)},
    symbolic x coords={ICD-IPD$_{PROFILE}$,ICD-IPD$_{SIM}$,ICD-IPD$_{DIF}$,ICD-IPD$_{WGCCA}$,ICD-IPD},
    x tick label style={rotate=25,anchor=east},
    xtick=data,
    ymajorgrids=true,
    my ybar legend,
    ]
\addplot [pattern=north east lines] coordinates {(ICD-IPD$_{PROFILE}$,84.69) (ICD-IPD$_{SIM}$,90.2) (ICD-IPD$_{DIF}$,71.29) (ICD-IPD$_{WGCCA}$,66.67) (ICD-IPD,94)};
\addplot [pattern=dots] coordinates {(ICD-IPD$_{PROFILE}$,26.31) (ICD-IPD$_{SIM}$,72.33) (ICD-IPD$_{DIF}$,24.2) (ICD-IPD$_{WGCCA}$,21.06) (ICD-IPD,79.29)};
\addplot [pattern=horizontal lines] coordinates {(ICD-IPD$_{PROFILE}$,40.12) (ICD-IPD$_{SIM}$,80.28) (ICD-IPD$_{DIF}$,36.12) (ICD-IPD$_{WGCCA}$,30.05) (ICD-IPD,86.01)};
\legend{Precision,Recall,F1-Score}
\end{axis}
\end{tikzpicture}
}}

\centering
\subfloat[][\textbf{50\%}]{\resizebox{0.35\textwidth}{!}{
\begin{tikzpicture} 
\begin{axis}[
    ybar,
    legend style={at={(0.5,-0.35)},anchor=north,legend columns=-1},
    width=11cm,
    height=6cm,
    ymin=20,
    ymax=90,
    ytick={20,30,40,50,60,70,80,90},
    enlargelimits=0.15,
    grid style=dotted,
    ylabel={\textbf{Performance} (\%)},
    symbolic x coords={ICD-IPD$_{PROFILE}$,ICD-IPD$_{SIM}$,ICD-IPD$_{DIF}$,ICD-IPD$_{WGCCA}$,ICD-IPD},
    x tick label style={rotate=25,anchor=east},
    xtick=data,
ymajorgrids=true,
    my ybar legend,
    ]
\addplot [pattern=north east lines] coordinates {(ICD-IPD$_{PROFILE}$,86.07) (ICD-IPD$_{SIM}$,90.48) (ICD-IPD$_{DIF}$,72.78) (ICD-IPD$_{WGCCA}$,67.15) (ICD-IPD,93.44)};
\addplot [pattern=dots] coordinates {(ICD-IPD$_{PROFILE}$,26.04) (ICD-IPD$_{SIM}$,74.76) (ICD-IPD$_{DIF}$,24.67) (ICD-IPD$_{WGCCA}$,21.81) (ICD-IPD,79.42)};
\addplot [pattern=horizontal lines] coordinates {(ICD-IPD$_{PROFILE}$,39.98) (ICD-IPD$_{SIM}$,81.87) (ICD-IPD$_{DIF}$,36.84) (ICD-IPD$_{WGCCA}$,32.93) (ICD-IPD,85.85)};
\legend{Precision,Recall,F1-Score}
\end{axis}
\end{tikzpicture}
}}

\subfloat[][\textbf{60\%}]{\resizebox{0.35\textwidth}{!}{
\begin{tikzpicture} 
\begin{axis}[
    ybar,
    legend style={at={(0.5,-0.35)},anchor=north,legend columns=-1},
    width=11cm,
    height=6cm,
    ymin=20,
    ymax=90,
    ytick={20,30,40,50,60,70,80,90},
    enlargelimits=0.15,
    grid style=dotted,
    ylabel={\textbf{Performance} (\%)},
    symbolic x coords={ICD-IPD$_{PROFILE}$,ICD-IPD$_{SIM}$,ICD-IPD$_{DIF}$,ICD-IPD$_{WGCCA}$,ICD-IPD},
    x tick label style={rotate=25,anchor=east},
    xtick=data,
    ymajorgrids=true,
    my ybar legend,
    ]
\addplot [pattern=north east lines] coordinates {(ICD-IPD$_{PROFILE}$,84.67) (ICD-IPD$_{SIM}$,90.16) (ICD-IPD$_{DIF}$,70.39) (ICD-IPD$_{WGCCA}$,66.33) (ICD-IPD,94.05)};
\addplot [pattern=dots] coordinates {(ICD-IPD$_{PROFILE}$,23.63) (ICD-IPD$_{SIM}$,72.54) (ICD-IPD$_{DIF}$,24.93) (ICD-IPD$_{WGCCA}$,20.166) (ICD-IPD,80.09)};
\addplot [pattern=horizontal lines] coordinates {(ICD-IPD$_{PROFILE}$,36.95) (ICD-IPD$_{SIM}$,80.39) (ICD-IPD$_{DIF}$,36.79) (ICD-IPD$_{WGCCA}$,30.92) (ICD-IPD,86.5)};
\legend{Precision,Recall,F1-Score}
\end{axis}
\end{tikzpicture}
}}
    \caption{Impact of the employed features of the proposed ICD-IPD.}
  \label{ablation_study}
\end{figure}
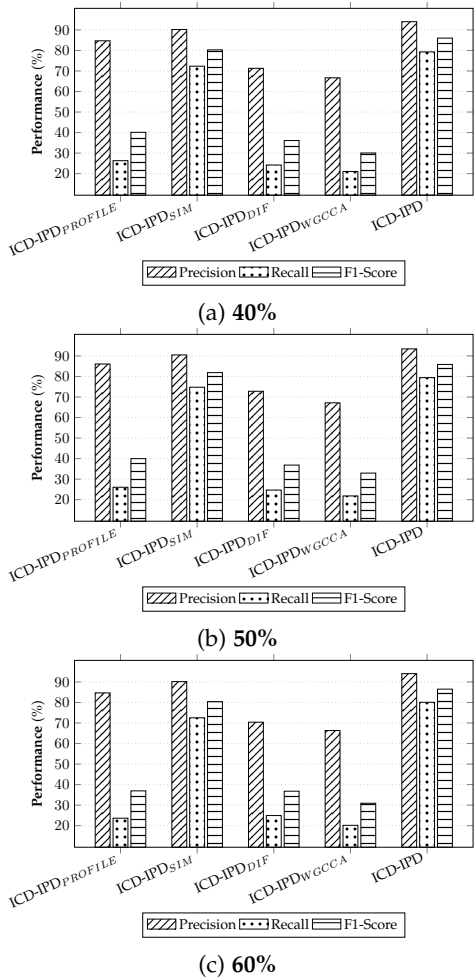
Figure \ref{ablation_study} shows the performance results of the ICD-IPD based on different variants when detecting cloned identities. The ICD-IPD based on all proposed features achieves better performance than the other variants on all evaluation metrics. We observed that ICD-IPD$_{SIM}$ performs better than the other variants for all different percentages. We also found that ICD-IPD$_{WGCCA}$ performs poorly compared to the other variants. We conclude that ICD-IPD$_{WGCCA}$ missing important features (e.g. similarity-based features) led to the aforementioned observation. The ICD-IPD aims to predict whether an account pair consists of cloned and its victim. Therefore,  the WGCCA-based feature cannot compare the account pair. On the other hand,  the ICD-IPD$_{SIM}$ achieved the best-performing result. The reason is that the similarity-based features compare the textual features, which can notably affect in identity cloning detection. The attacker mostly needs to mimic the textual features (e.g. screen name, description) of the victim to convince the victim. Overall, the performance results indicated that the employed features together in the ICD-IPD provide the best performance results.

\section{Conclusion and Future Work}\label{concl}
This paper proposes a novel identity cloning detection approach in the face of incomplete non-privacy-sensitive profile data, named ICD-IPD. 
ICD-IPD 
was evaluated against the existing state-of-the-art identity cloning detection approaches and other machine or deep learning models atop a real-world dataset. The results of our extensive evaluation show that ICD-IPD outperforms all the 
state-of-the-art identity cloning detection as well as other machine and deep learning approaches compared.
 {Our future work will aim to explore additional datasets as they become available or develop methods to augment our existing data to further validate our approach. In addition, we plan to implement the proposed approach in a real-world setting and conduct manual validation of detected cloned accounts to assess its practical effectiveness. }


\section*{Acknowledgement}
This work is funded by the Australian Research Council under Grant No. DP220101823.


%





\ifCLASSOPTIONcaptionsoff
  \newpage
\fi



\bibliographystyle{IEEEtran}
\bibliography{ref.bib}
%



%
\begin{IEEEbiography}[{\includegraphics[width=1in,height=1.25in,clip,keepaspectratio]{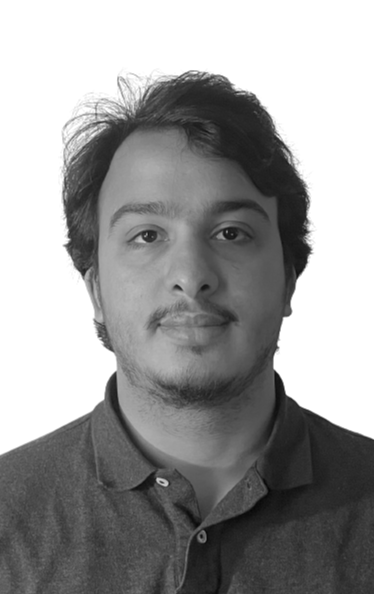}}]{Ahmed Alharbi} received the PhD in Computer Science from RMIT University, Australia in 2023. He is currently an assistant professor at the College of Computer Science and Engineering in Taibah University, Medina, Saudi Arab. His publications appear in \textit{ACM Computing Surveys}, \textit{ICSOC}, \textit{ICWS}, etc. His research interests include identity cloning detection and application of machine learning in social networks.
\end{IEEEbiography}
\begin{IEEEbiography}[{\includegraphics[width=1in,height=1.25in,clip,keepaspectratio]{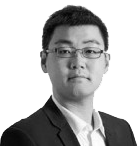}}]{Hai Dong} is currently a senior lecturer at School of Computing Technologies in RMIT University, Melbourne, Australia. 
He received a PhD from Curtin University, Australia and a Bachelor's degree from Northeastern University, China.
His research interests include Service-Oriented Computing, Distributed Systems, Cyber Security, Machine Learning and Data Science. He is a senior member of the IEEE.
\end{IEEEbiography}
\begin{IEEEbiography}[{\includegraphics[width=1in,height=1.25in,clip,keepaspectratio]{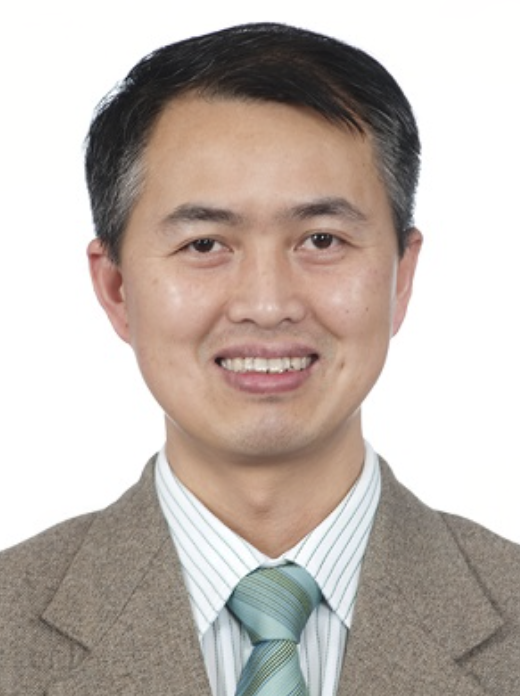}}]{Xun Yi} is currently a Professor in Cyber Security with School of Computing Technologies, RMIT University, Australia. His research interests include Cloud and IoT Security and Privacy, Distributed System Security, Blockchain Applications, and Applied Cryptography. 
Currently, he is an Associate Editor with IEEE Transactions on Knowledge and Data Engineering. He has been an ARC College Expert from 2017 to 2019.
\end{IEEEbiography}
\begin{IEEEbiography}[{\includegraphics[width=1in,height=1.25in,clip,keepaspectratio]{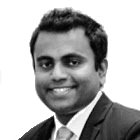}}]{Prabath Abeysekara} received his B.Sc. (Hons) of Engineering degree from the University of Moratuwa, Sri Lanka in 2010 and a PhD in Computer Science at RMIT University, Melbourne, Australia in 2022. His primary research interests include Machine Learning, Cyber Security and Distributed Computing. 
\end{IEEEbiography}







\end{document}